\documentclass[aps,prb,twocolumn]{revtex4}
\usepackage{bm,color,amsmath,amssymb,mathrsfs,latexsym,graphicx,psfrag}

\newcommand{\bk}{{\mathbf k}}
\newcommand{\bq}{{\mathbf q}}
\newcommand{\br}{{\mathbf{r}}}
\newcommand{\be}{\begin{equation}}
\newcommand{\ee}{\end{equation}}

\def\be{\begin{equation}}
\def\ee{\end{equation}}
\def\bea{\begin{eqnarray}}
\def\eea{\end{eqnarray}}

\begin{document}

\title{Large Chern Number Quantum Anomalous Hall Effect In \\Thin-film Topological Crystalline Insulators}
\author{Chen Fang$^{1,2}$, Matthew J. Gilbert$^{3,4}$, B Andrei Bernevig$^2$}
\date{\today}
\affiliation{$^{1}$Department of Physics, University of Illinois, Urbana IL 61801-3080}
\affiliation{$^2$Department of Physics, Princeton University, Princeton NJ 08544}
\affiliation{$^3$Department of Electrical and Computer Engineering, University of Illinois, Urbana IL 61801}
\affiliation{$^4$Micro and Nanotechnology Laboratory, University of Illinois, Urbana IL 61801}
\begin{abstract}{Quantum anomalous Hall (QAH) insulators are two-dimensional (2D) insulating states exhibiting properties similar to those of quantum Hall states but without external magnetic field. They have quantized Hall conductance $\sigma^H=Ce^2/h$, where integer $C$ is called the Chern number, and represents the number of gapless edge modes. Recent experiments demonstrated that chromium doped thin-film (Bi,Sb)$_2$Te$_3$  is a QAH insulator with Chern number $C=\pm1$. Here we theoretically predict that thin-film topological crystalline insulators (TCI) can host various QAH phases, when doped by ferromagnetically ordered dopants. Any Chern number between $\pm4$ can, in principle, be reached as a result of the interplay between (a) the induced Zeeman field, depending on the magnetic doping concentration, (b) the structural distortion, either intrinsic or induced by a piezoelectric material through proximity effect and (c) the thickness of the thin film. The tunable Chern numbers found in TCI possess significant potential for ultra-low power information processing applications.}
\end{abstract}
\maketitle

A quantum anomalous Hall state is a 2D topological insulating state that has quantized Hall conductance in the form of $Ce^2/h$ where $C$ is an integer, and possesses $|C|$ gapless edge modes along any 1D edge. These properties are shared by the well-known quantum Hall states\cite{Klitzing}. Nevertheless, there is no external magnetic field in a QAH state, which makes it `anomalous'. Hence, the nontrivial topology in QAH does not come from the topology of the Landau levels, but rises from the band structure of electrons coherently coupled to certain magnetic orders, e.g., spin orders and orbital current orders. The first theoretical model that shows this phase is given in Ref.[\onlinecite{Haldane1988}], which is followed by other models and experimental proposals in various systems\cite{Onoda2003,Qi2008,Yu2010,Nomura2011,Jiang2012,Wang2013}. Very recently, experimentalists have adapted one of the proposals and realized a QAH state with $|C|=1$ in chromium doped thin-film (Bi,Sb)$_2$Te$_3$, which is a 3D topological insulator (TI)\cite{Chang2013}.

We first recapitulate the basic idea underlying the realization of QAH insulators with $|C|=1$ in a thin-film 3D topological insulator\cite{Qi2008,Yu2010,Nomura2011,Jiang2012}. Each surface of a 3D TI is a gapless 2D Dirac spin-split semi-metal\cite{hasan2010,qi2011rev}, as opposed to spin-degenerate Dirac semi-metals such as grapehene. The surface is spin-split except at the Dirac point where double-degeneracy is protected by time-reversal symmetry, and spectral flow into the bulk conduction and valence bands occurs away from the Dirac point. Upon the application of a Zeeman field along the perpendicular direction, induced by ferromagnetic dopants, a gap is opened at the Dirac point, giving rise to a massive Dirac cone. Such a massive Dirac cone has been well known to contribute Hall conductance of $\pm{}e^2/2h$\cite{Qi2008,Bernevig2013}, or, a Chern number of $\pm{1/2}$. Moreover, since a thin film has two surfaces (top and bottom), the total Chern number is $\pm1$. An identical effect would take place in bulk samples - thin films are being used here only because they allow tuning of the Fermi level in the gap by gating. Here we use a symmetry-based analysis to show that the topological crystalline insulators\cite{Fu2011,Hsieh2012,Xu2012,Dziawa2012,Tanaka2012,Fang2012a,Wang2013a,Liu2013,Okada2013} [such as (Pb,Sn)(Te,Se)] are much richer compounds to explore QAH physics. As thin films of (Pb,Sn)(Te,Se) have already been grown\cite{Elleman1948,Bylander1966,Taskin2013} and various magnetic dopants have been successfully doped\cite{Mathur1970,Inoue1976,Nielsen2012}, we believe our proposal is experimentally realizable. The existence of such a widely tunable topological phase transition in the TCI class of materials may form the basis for new types of information processing devices which consume much less power compared to current technology.

\section{Results}

\subsection{Unperturbed Hamiltonian on the $(001)$-surface}

\begin{figure}
\includegraphics[width=8cm]{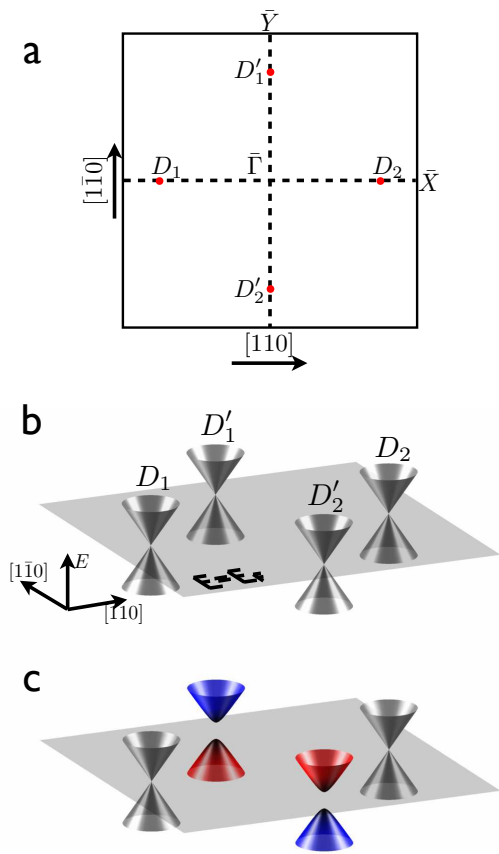}
\caption{(\textbf{a}) The surface Brillouin zone centered at $\bar\Gamma$ and bounded by $\bar{X},\bar{Y}$, which is symmetric under 90-degree rotations about the vertical line through the center, and mirror reflections about the two dotted lines. The positions of the Dirac points are marked. (\textbf{b}) The schematics of the dispersion of the four Dirac cones on the $(001)$-plane in the SBZ. The middle plane is the $E=E_F$ plane passing through the four Dirac points at exact half filling. (\textbf{c}) The proposed surface dispersion of the rhomboderal phase with two massive and two massless Dirac cones, where a red/blue cone contributes a fractional Chern number of $+1/2$/$-1/2$, respectively.}
\label{fig:BZ}
\end{figure}

Consider the symmetries of such a thin film. (Pb,Sn)(Te,Se) crystalizes into a face-centered-cubic lattice with point group $O_h$. Below a critical temperature, depending on composition, the cubic symmetry spontaneously breaks into either rhombohedral or orthorhombic symmetries, resulting in a small lattice distortion. Here we assume that the lattice has cubic symmetry and treat the small distortion as perturbative strain. The thin-film sample is terminated on the $(001)$-plane, where $O_h$ reduces to 2D point group $C_{4v}$. The bulk system also has time-reversal symmetry and inversion symmetry, which relates the top and the bottom surfaces in the absence of asymmetric surface terminations. The in-plane translational symmetry allows the definition of the surface Brillouin zone (SBZ), which is centered at $\bar\Gamma$ and bounded by $\bar{X}$ along the $[110]$-direction and $\bar{Y}$ along $[1\bar{1}0]$-direction [Figure \ref{fig:BZ}(a)]. Four Dirac points close to the Fermi energy have been observed in experiments\cite{Xu2012,Tanaka2012,Dziawa2012}. Two Dirac points, denoted by $D_{1,2}$, are located along $\bar\Gamma\bar{X}$, close to and symmetric about $\bar{X}$; two others, denoted by $D_{1',2'}$, are located along $\bar\Gamma\bar{Y}$, close to and symmetric about $\bar{Y}$. The band dispersion around any of the four Dirac points is linear in all directions to first order, resulting in four copies of a spin-split Dirac semi-metal, related to each other by 90-degree rotations [Figure \ref{fig:BZ}(b)]. Recently, scanning tunneling spectroscopy (STM) measurements suggest\cite{Okada2013} that in the rhombohedral phase, two of the four Dirac points are gapped [Figure \ref{fig:BZ}(c)].

We assume that the Fermi level is exactly at the Dirac point energy. While this is not true in bulk samples due to intrinsic impurity doping, in thin-film samples the Fermi level can be tuned anywhere in the bulk gap. Since the change in the Chern number only depends on the electronic states near the gap-closing points, i.e., the four Dirac points, we start by deriving the effective theories for each Dirac cone and then consider their coupling to gap-opening perturbations. The minimal model for each Dirac cone $h_{i=1,2,1',2'}(\bq)$, where $\bq=\bk-\mathbf{D}_i$, is a two-band $k\cdot{p}$ model, due to the double-degeneracy at $D_i$. The form of $h_{i}$ is determined by how the doublet at $D_i$ transforms under the little group at $D_i$, i.e., a subgroup of the full symmetry group which leaves $D_i$ invariant. For  example, consider $D_1$: the little group is generated by the mirror reflection about the $(1\bar10)$-plane, denoted by $M_{1\bar10}$ and a combined operation of a 180-degree rotation about the $[001]$-direction followed by time-reversal, denoted by $C_{2T}$. This little group has only one 2D irreducible representation (see Sec.\ref{sec:littlegroup}): $M_{1\bar10}=i\sigma_y$ and $C_{2T}=K\sigma_x$, where $K$ means complex conjugation, and $\sigma_{x,y,z}$ are Pauli matrices. It restricts $h_1(\bq)$ to the form
\bea\label{eq:h_1}
h_1(\bq)=v_0q_1I_{2\times2}+v_1q_1\sigma_y+v_2q_2\sigma_x
\eea
up to the first order of $|q|$. $\bq$ decomposes into two components $q_1=\bq\cdot\hat{e}_{110}$ and $q_2=\bq\cdot\hat{e}_{1\bar10}$, where $\hat{e}_{mnl}$ is the unit vector along the $[mnl]$-direction.
The parameters $v_{0,1,2}$ can be fixed by matching the dispersion of equation (\ref{eq:h_1}), $E(\bq)=v_0q_1\pm\sqrt{v_1^2q_1^2+v_2^2q_2^2}$ to the measured Fermi velocities along $[110]$- and $[1\bar10]$-directions [$(v_0,|v_1|,|v_2|)\sim(0,1.1,2.8)$eV\AA]. The Dirac cones centered at $D_{2,1',2'}$ can be related to the cone centered at $D_1$ by $C_4$ symmetry. This automatically gives the effective theories of the other Dirac cones: $h_2(q_1,q_2)=h_1(-q_1,-q_2)$, $h_{1'}(q_1,q_2)=h_1(-q_2,q_1)$ and $h_{2'}(q_1,q_2)=h_1(q_2,-q_1)$ (see Sec.\ref{sec:fourDirac} for a formal proof).

\begin{figure*}
\includegraphics[width=18cm]{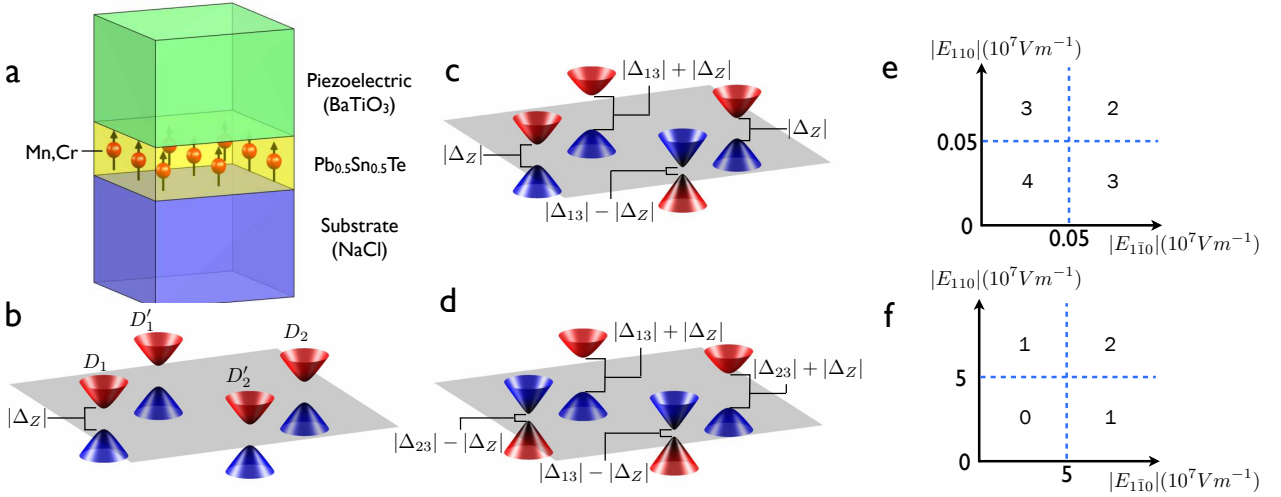}
\caption{(\textbf{a}) The schematic of a thin-film Pb$_{0.5}$Sn$_{0.5}$Te grown on a substrate, capped by a piezoelectric. (\textbf{b-d}) The schematic dispersions of the gapped Dirac cones on the top surface in the presence of uniform Zeeman field and strains, corresponding to the parameters $\Delta_Z>|\Delta_{13}|=|\Delta_{23}|=0$, $|\Delta_{13}|>\Delta_Z>|\Delta_{23}|=0$, and $0<\Delta_Z<|\Delta_{13}|=|\Delta_{23}|$, respectively. (\textbf{e}) The Chern number of the proposed system in the thick limit (sample thickness $>20$nm) plotted against the transverse electric fields applied on the piezoelectric. (\textbf{f}) The Chern number of the system with thickness of $5\sim10$nm.}
\label{fig:phases}
\end{figure*}

\subsection{The effect of induced Zeeman field}

We assume a Zeeman field in the sample along $[001]$-direction, induced by ferromagnetically ordered dopants. In order to couple this field to the electrons in the $k\cdot{p}$ models, we add an additional term $\delta{H}^Z_i$ to $h_{i=1,2,3,4}(\bq)$ and note the following facts: (i) magnetization along $[001]$-direction changes sign under both $M_{1\bar10}$ and $C_{2T}$ and (ii) it is invariant under 90-degree rotations about $[001]$-direction. Using these facts, we have:
\bea\nonumber
\delta{H}^Z_i\equiv\delta{H}^Z=\Delta_Z\sigma_z+O(|q|),
\eea
where $|\Delta_Z|$ is the field strength of the Zeeman field, which is proportional to the Curie temperature, $T_c$, of the ferromagnetism. The sign of $\Delta_Z$ depends on the direction of the magnetization. The Hamiltonian for each cone with the induced Zeeman field is $h_i(\bq)+\delta{H}^Z$, which has a gap of size $|\Delta_Z|$ at each Dirac point [see Figure \ref{fig:phases}(b)]. 

\subsection{The effect of intrinsic and applied strain}
\label{sec:strain}
\begin{table}
\label{tab:epsilon}
\begin{center}
\begin{tabular}{|c|c|c|c|c|c|c|}
\hline
 & $\epsilon_{11}$ & $\epsilon_{22}$ & $\epsilon_{33}$ & $\epsilon_{12}$ & $\epsilon_{13}$ & $\epsilon_{23}$\\
\hline
\hline
$C_{2T}$ & $+$ & $+$ & $+$ & $+$ & $-$ & $-$\\
\hline
$M_{1\bar10}$ & $+$ & $+$ & $+$ & $-$ & $+$ & $-$\\
\hline
$M_{110}$ & $+$ & $+$ & $+$ & $-$ & $-$ & $+$\\
\hline
$C_4$ & $\epsilon_{22}$ & $\epsilon_{11}$ & $+$ & $-$ & $\epsilon_{23}$ & $-\epsilon_{13}$\\
\hline
$\delta{H}^S_1$ & $\sigma_y$ & $\sigma_y$ & $\sigma_y$ & $\sigma_x$ & $\sigma_z$ & $0$\\
\hline
$\delta{H}^S_2$ & $\sigma_y$ & $\sigma_y$ & $\sigma_y$ & $\sigma_x$ & $-\sigma_z$ & $0$\\
\hline
$\delta{H}^S_{1'}$ & $\sigma_y$ & $\sigma_y$ & $\sigma_y$ & $-\sigma_x$ & $0$ & $\sigma_z$\\
\hline
$\delta{H}^S_{2'}$ & $\sigma_y$ & $\sigma_y$ & $\sigma_y$ & $-\sigma_x$ & $0$ & $-\sigma_z$\\
\hline
\end{tabular}
\caption{First four rows show the transformation properties of each tensor component under the symmetry group $C_{4v}\otimes{T}$, where $\pm$ means invariant or inverted. The last four rows show which Pauli matrix is coupled to each component, to the zeroth order, in the effective theory for each Dirac cone.}
\end{center}
\end{table}

Now we consider the effect of intrinsic and external strains. Depending on Sn and Se concentration, the cubic lattice can have spontaneous distortions into either rhombohedral or the rhombohedral symmetries. One may also cap the top surface of the film with a piezoelectric material such as BaTiO$_3$, to control the strain on the top surface. A general strain tensor is given by a symmetric matrix $\epsilon_{ij}$ where $i,j=1,2,3$, written in the frame spanned by $(\hat{e}_{110},\hat{e}_{1\bar10},\hat{e}_{001})$. In order to represent couplings to the strain tensor in the $k\cdot{p}$ models, we need to determine the transform of each component under the symmetry group $C_{4v}$ and time-reversal (Table I). Using these relations, we obtain the following strain induced terms for the four Dirac cones, to the zeroth order of $|q|$:
\bea\nonumber
\delta{H}_{1/2}^S&=&(\lambda_{11}\epsilon_{11}+\lambda_{22}\epsilon_{22}+\lambda_{33}\epsilon_{33})\sigma_y\\
\nonumber&+&\lambda_{12}\epsilon_{12}\sigma_x\pm\lambda_{23}\epsilon_{23}\sigma_z,\\
\nonumber\delta{H}_{1'/2'}^S&=&(\lambda_{11}\epsilon_{22}+\lambda_{22}\epsilon_{11}+\lambda_{33}\epsilon_{33})\sigma_y\\
\nonumber&-&\lambda_{12}\epsilon_{12}\sigma_x\pm\lambda_{13}\epsilon_{13}\sigma_z,
\eea
where $\lambda_{ij}$ are electro-phonon couplings.

Consider the full Hamiltonian for each Dirac cone under both Zeeman field and strain, $H_i=h_i+\delta{H}_i^Z+\delta{H}_i^S$. In $H_i$, only terms proportional to $\sigma_z$ open gaps in the spectrum while others move the position of the Dirac point $D_i$. The gap at each $D_i$, i.e., the coefficient before the $\sigma_z$ term in the Hamiltonians, denoted below by $\Delta_i$, is:
\bea\nonumber
\Delta_{1,2}=\Delta_Z\pm\Delta_{23},\\
\nonumber
\Delta_{1',2'}=\Delta_Z\pm\Delta_{13},
\eea
where we have defined $\Delta_{13/23}\equiv\lambda_{23/13}\epsilon_{23/13}$. Each gapped Dirac cone contributes
\bea\label{eq:hall1}
\sigma^H_i=-\textrm{sign}(v_1v_2\Delta_i)e^2/(2h)
\eea
to the Hall conductance (see Sec.\ref{sec:Chern} for formal proof).

\subsection{The effect of finite thickness}

\begin{figure}
\includegraphics[width=8cm]{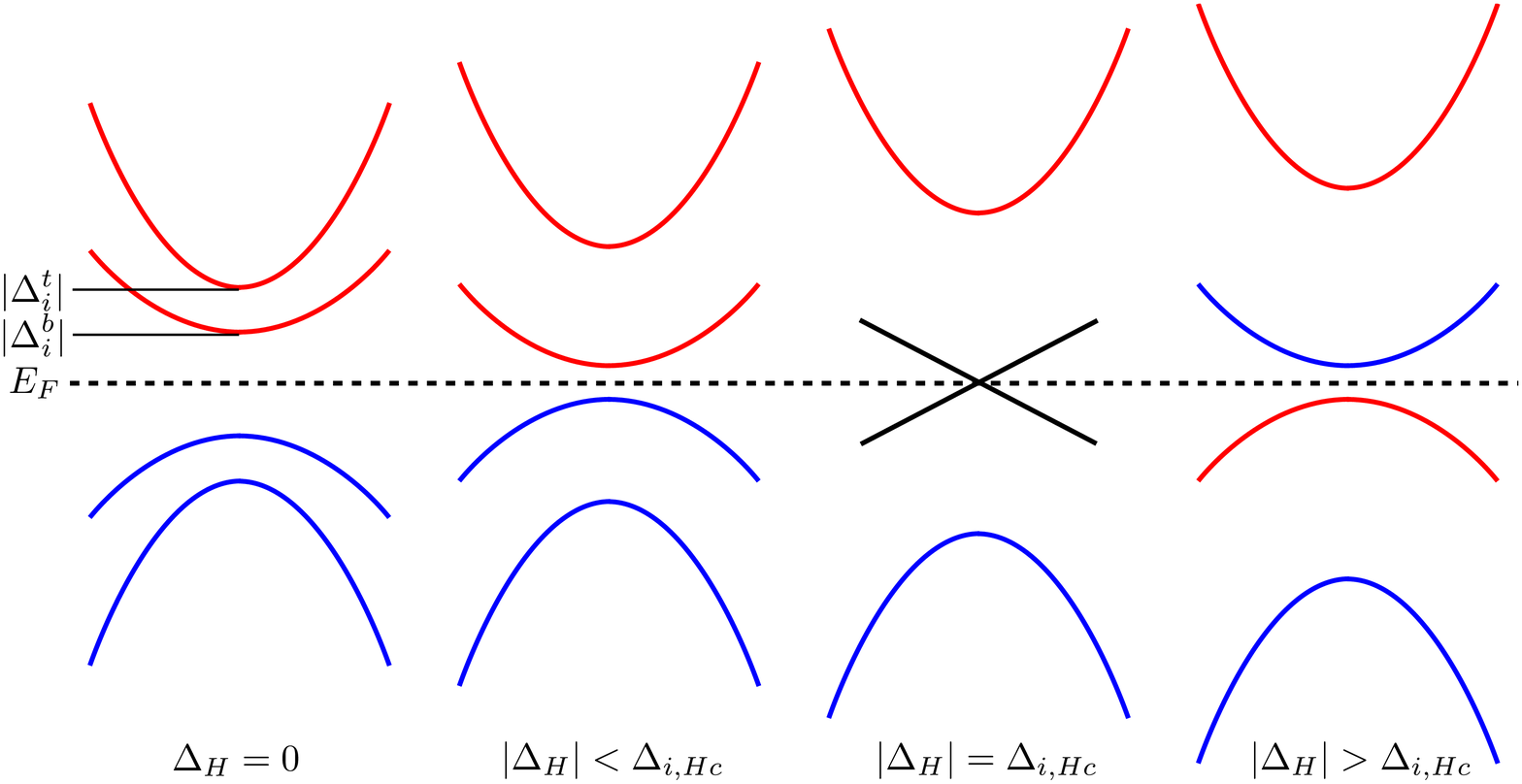}
\caption{The figures shows the quantum phase transition happening at the Dirac cone centered at $D_i$ which is induced by increasing the hybridization gap, $\Delta_H$, between the top and the bottom surfaces. Here we start from a cone with $\Delta^t_i>\Delta^b_i>0$ at $\Delta_H=0$. Red/blue means the represented cone contributes $\pm1/2$ to the Chern number, respectively.}
\label{fig:Delta_H}
\end{figure}

We have so far assumed that the top and the bottom surfaces are isolated from each other, and hence the total Hall conductance is 
\bea\label{eq:hall_thick}
\sigma^H=\sum_{i=1,2,1',2'}\sigma_i^{H,t}+\sigma_i^{H,b},
\eea
where superscript $t/b$ denotes the top/bottom surface.
When the thickness is comparable to the decay length of the surface states, the hybridization gap between the two surfaces, denoted by $\Delta_H$, becomes significant, and the total Hall conductance is generically not given by equation (\ref{eq:hall_thick}). Diagonalizing each $k\cdot{p}$ Hamiltonian with hybridization (see Sec.\ref{sec:dispersion} for the explicit forms of the band dispersion)
\bea\label{eq:hybridization}
\tilde{H}_i=\left(\begin{matrix}
H^t_i & \Delta_HI_{2\times2}\\
\Delta_HI_{2\times2} & H^b_i
\end{matrix}\right),
\eea
we have two scenarios. (i) If $\textrm{sign}(\Delta_i^t)=\textrm{sign}(\Delta_i^b)$ (where $\Delta^{t/b}$ denotes the gap at top/bottom surface), as $\Delta_H$ increases, the gap at $D_i$ closes at $|\Delta_H|=\Delta_{i,Hc}\equiv\sqrt{|\Delta^t_i\Delta^b_i|}$ and reverses [see Figure \ref{fig:Delta_H}], and at $|\Delta_H|>\Delta_{i,Hc}$, the total contribution to $\sigma^H$ vanishes; (ii) if $\textrm{sign}(\Delta_i^t)=-\textrm{sign}(\Delta_i^b)$, there is no quantum phase transition as $\Delta_H$ increases, and the total contribution to Hall conductance stays at zero. The complete expression for the Hall conductance is therefore
\bea\label{eq:hall_full}
\sigma^H=\sum_{i=1,2,1',2'}(\sigma_i^{H,t}+\sigma_i^{H,b})\theta(\Delta_{i,Hc}-|\Delta_H|),
\eea
where $\theta(x)$ is the Heaviside step function.

\subsection{Proposals of materials and experiments}

Depending on the parameter set of $\{\Delta_Z,\Delta^{t/b}_{13,23},\Delta_H\}$, the Chern number of the system takes each integer from $-4$ to $+4$. In a realistic system, however, not all parameters are easily tunable, so the range of the Chern number is generically restricted. We propose a system shown in Figure \ref{fig:phases}(a): a thin-film Pb$_{0.5}$Sn$_{0.5}$Te doped with Mn or Cr, grown on a substrate, e.g., NaCl or KCl, with its top surface deposited with piezoelectric crystal such as BaTiO$_3$. Below $T\sim10$K, the (Cr,Mn) moments develop ferromagnetism, inducing a small Zeeman gap $\Delta_Z\sim1$meV in the sample\cite{Mathur1970}. The external strain on the top surface may be tuned by the piezoelectric. Assuming that the strain in BaTiO$_3$ be completely transferred to the top surface of the film, we estimate that\cite{Berlincourt1958,Bierly1963,Littlewood2010} the $|\Delta^{t}_{13}|=2\times10^{-6}E_{1\bar10}$meV$\cdot$m$\cdot$V$^{-1}$ and $|\Delta^{t}_{23}|=2\times10^{-6}E_{110}$meV$\cdot$m$\cdot$V$^{-1}$. Since the sample with such composition has zero or negligible intrinsic distortion at low temperatures, $\Delta_{13,23}^b=0$. In the thick limit ($d>20$nm), $\Delta_H\ll1$meV and is negligible\cite{Lin}. From equation (\ref{eq:hall1}), the bottom surface always contributes $C=2$. There are three possible scenarios for the top surface, resulting in $\sigma^{H,t}=2,1,0$ respectively: (i) $|\Delta^t_{13,23}|<|\Delta_Z|$, (ii) $|\Delta^t_{23}|<|\Delta_Z|<|\Delta^t_{13}|$ and (iii) $|\Delta_Z|<|\Delta^t_{13,23}|$, where we have assumed $|\Delta^t_{13}|>\Delta^t_{23}|$ without loss of generality. The dispersion of the four gapped cones for the three scenarios are plotted in Figure \ref{fig:phases}(b-d). The total Chern number can thus be tuned between $2$, $3$ and $4$, plotted against $E_{1\bar10}$ and $E_{110}$ in Figure \ref{fig:phases}(e). In a thiner film with thickness $d=5\sim10$nm, the hybridization gap is $|\Delta_H|=5\sim15$meV\cite{Liu2013b}, from which we take $\Delta_H=10$meV as a typical value and we plot the Chern number against $E_{1\bar10}$ and $E_{110}$ in Figure \ref{fig:phases}(f). From this Figure, we see that around the critical field strength $|E_{1\bar10}|=|E_{110}|=5\times10^7$Vm$^{-1}$, the Chern number can be electrically tuned to $0$, $1$ or $2$. If the length and width of the sample are both $100$nm, this means that the Chern number can be tuned by varying $V_{1\bar10,110}$ within 10mV. The ability to tune the topological phase transition with such a small electric field offers hope that such a logic devices based on piezoelectric deformation of a TCI could possess on/off ratios and sub-threshold slopes which far exceed current logic device technologies.

\section{Discussion}

In the derivation of the main results, we have ignored physical factors of (i) the impurities and (ii) the electron-electron interaction. The mirror Chern number of a TCI is only well defined in the presence of mirror planes. In a system with a random impurity configuration, mirror symmetries are broken and the mirror Chern number is not a good quantum number, and consistently, the gapless modes at the Dirac points are gapped by impurity scattering. This mirror symmetry breaking by impurity has, however, no effect on the Chern number in a ferromagnetically doped system, as long as the intensity of the random potential is much smaller compared with the Zeeman gap. This is because the Chern number, unlike the mirror Chern number, does not presume any symmetry, as long as the surface is gapped. Weak interactions smaller than the Zeeman gap do not have any effect on the quantized Hall conductance either, because the Chern number is also a good quantum number of an interacting gapped 2D system\cite{qi2006,Kohmoto1985}.

It is also interesting to discuss other surface terminations besides the $(001)$-surface. On the $(110)$-surface of SnTe, first principles calculation\cite{Liu2013} shows that there are two Dirac cones centered at two Dirac points that are close to and symmetric about $\bar{X}$ along $\bar\Gamma\bar{X}$ in the surface BZ. The two Dirac points are protected by the $(1\bar10)$ mirror plane and have equal energy due to the $(001)$ mirror plane. A Zeeman field along $[110]$ gaps both Dirac points and results in a QAH phase with Chern number of $\pm2$. A strain along $[1\bar11]$-direction breaks both the $(1\bar10)$ and the $(001)$ mirror planes, opening two gaps of opposite signs at the two Dirac points. When both the strain and the Zeeman field are present, a discussion similar to the one given in Sec.\ref{sec:strain} shows that the Chern number can be either $\pm1$ or $\pm2$. On the $(111)$-surface, there are four Dirac cones centered at $\bar\Gamma$ and three $\bar{M}$'s. The three Dirac points at $\bar{M}$ have the same energy due to the threefold rotation symmetry about the $[111]$-axis, while the one at $\bar\Gamma$ generically has a different energy. This energy difference among the Dirac points, which has been measured to be $\sim40$meV in Ref.[\onlinecite{Taskin2013}], makes it hard to have a fully gapped surface using an induced Zeeman field, because the Zeeman gap is generically much smaller than $40$meV. Therefore, an insulator with quantized Hall conductance on the $(111)$-surface is not possible using the current scheme.

\section{Methods} 
\subsection{Derivation of $h_1(\bq)$ using the little group at $D_1$}
\label{sec:littlegroup}
The full symmetry group of the thin film in the absence of applied fields is $D_{4h}\otimes{T}$. The little group at a Dirac point $D_i$ is the subgroup of all operations that leave $D_i$ invariant. The little group therefore consists of a mirror plane that passes $\bar\Gamma{}D_i$, $C_{2T}$ and their combinations. Taking $D_1$ as example, the little group is generated by $M_{1\bar10}$ and $C_{2T}$. In a general spin-$1/2$ system we have: $M^2_{1\bar10}=C_2^2=T^2=-1$ and $\{M_{1\bar10},C_2\}=[M^2_{1\bar10},T]=[C_2,T]=0$, where $C_2$ is the 180-rotation about $[001]$-direction. Therefore the two generators satisfy (i) $M_{1\bar10}^2=-C_{2T}^2=-1$ (ii) $\{M_{1\bar10},C_{2T}\}=0$. There is only one 2D irreducible representation up to a basis rotation: $M_{1\bar10}=i\sigma_y$ and $C_{2T}=K\sigma_x$. Physically, $M_{1\bar10}$ relates the Hamiltonian $h_1(q_1,q_2)$ to $h_1(q_1,-q_2)$ and $C_{2T}$ commutes with $h_1(\bq)$; or mathematically, $M_{1\bar10}h_1(q_1,q_2)M^{-1}_{1\bar10}=h_1(q_1,-q_2)$ and $[C_{2T},h_1(\bq)]=0$. The irreducible representation of the little group along with the symmetry constraints determine the form of $h_1(\bq)$ shown in equation (\ref{eq:h_1}).

In general, the $k\cdot{p}$ model is given by
\bea
h_1(\bq)=d_0(\bq)I_{2\times2}+d_x(\bq)\sigma_x+d_y(\bq)\sigma_y+d_z(\bq)\sigma_z,
\eea
which must satisfy the symmetry constraints:
\bea
M_{1\bar10}h_1(q_1,q_2)M^{-1}_{1\bar10}&=&h_1(q_1,-q_2),\\
\nonumber
[C_{2T},h_1(\bq)]=0.
\eea
These symmetry constraints give that (1) $d_{0,y}$ is even under $q_2\rightarrow-q_2$, (2) $d_x$ is odd under $q_2\rightarrow-q_2$ and (3) $d_z=0$ to arbitrary order. We expand them to the second order in $|q|$:
\bea
d_0(q_1,q_2)&=&v_0q_1+\frac{q_1^2}{2m_1}+\frac{q_2^2}{2m_2},\\
d_x(q_1,q_2)&=&v_2q_2+\frac{q_1q_2}{2m_3},\\
d_y(q_1,q_2)&=&v_1q_1+\frac{q_1^2}{2m_4}+\frac{q_2^2}{2m_5}.
\eea
These terms make the dispersion deviate from perfectly linear and may be understood as the `warping' terms; they also make corrections to the wave functions at each $\bq$. It should be noted that $d_z=0$ holds up to arbitrary orders and this means there is no out-of-plain pseudo-spin component at any $\bq$. While including higher order terms explains the shape-changing of the equal energy contours from perfect ellipsoids, the Lifshitz transition cannot be described in the framework of any two-band theory. To do so, the model must be extended a four-band one, in order to account for the hybridization between nearest cones, as discussed in Ref.[\onlinecite{Fang2012a}].

\subsection{Relating the four Dirac cones by $C_4$ symmetry}
\label{sec:fourDirac}
In the main text, we mention that by 90-degree rotations the effective theories for the four cones can be related. This is an intuitive statement yet to be made precise. In fact, $k\cdot{p}$ theories are always written with respect to a chosen basis, which is our case is furnished by (the periodic part of) the two Bloch states that are degenerate at the Dirac point. Due to the degeneracy, there is a gauge degree of freedom in the choice. Here the choice is made by fixing the little group representation at $D_1$: $M_{1\bar10}=i\sigma_y$ and $C_{2T}=K\sigma_x$. If we denote the two basis states by $|u_{1\uparrow}\rangle$ and $|u_{1\downarrow}\rangle$, we then fix the bases at $D_{2,1',2'}$ to be $\{|u_{2\uparrow}\rangle,|u_{2\downarrow}\rangle\}=\{\tilde C^2_4|u_{1\uparrow}\rangle,\tilde C^2_4|u_{1\downarrow}\rangle\}$, $\{|u_{1'\uparrow}\rangle,|u_{1'\downarrow}\rangle\}=\{\tilde C_4|u_{1\uparrow}\rangle,\tilde C_4|u_{1\downarrow}\rangle\}$ and $\{|u_{2'\uparrow}\rangle,|u_{2'\downarrow}\rangle\}=\{\tilde C^3_4|u_{1\uparrow}\rangle,\tilde C^3_4|u_{1\downarrow}\rangle\}$, respectively. Mark that here $\tilde C_4$ is the matrix representing the 90-degree rotation in both orbital space (including spin). Defining the Bloch wave function at $\mathbf{D}_i+\bq$ as $|\psi_{i\uparrow/\downarrow}(\bq)\rangle=e^{i(\mathbf{D_i}+\bq)\cdot\br}|u_{i\uparrow/\downarrow}\rangle$, it is easy to check that $|\psi_{2}(\bq)\rangle=\hat{C}^2_4|\psi_1(-\bq)\rangle$, $\hat{C}_4|\psi_{1'}(\bq)\rangle=\hat{C}_4|\psi_1(-q_2,q_1)\rangle$ and $|\psi_{2'}(\bq)\rangle=\hat{C}_4^3|\psi_1(q_2,-q_1)\rangle$. Here $\hat{C}_4$ is the single particle operator acting in the Hilbert space, which is the combination of the orbital rotation $\tilde{C}_4$ plus rotation $(x,y)\rightarrow(-y,x)$, where $(x,y)$ is a lattice point and the rotation center is also placed at a lattice point. The full single Hamiltonian, projected to the states at the vicinities of the four Dirac points, is given by
\bea
\hat{H}=\sum_{\bq,i=1,2,1',2',\alpha,\beta=\uparrow,\downarrow}(h_i(\bq))^{\alpha\beta}|\psi_{i\alpha}(\bq)\rangle\langle\psi_{i\beta}(\bq)|.
\eea
$C_4$ symmetry implies $[\hat{C}_4,\hat{H}]=0$, which immediately leads to $h_2(q_1,q_2)=h_1(-q_1,-q_2)$, $h_{1'}(q_1,q_2)=h_1(-q_2,q_1)$ and $h_{2'}(q_1,q_2)=h_1(q_2,-q_1)$, confirming the intuitive relations appearing in the main text.

\subsection{Calculation of the Chern number of the top/bottom surface}
\label{sec:Chern}
In the text we refer to the Chern number contributed by one massive Dirac cone, which is not mathematically well-defined. In fact, the integrated Berry's curvature of a gapped Dirac cone is non-quantized in any finite $\bk$-space, hence possesses no well-defined Chern number. The Chern number of a whole 2D surface (top surface for example) is, however, a well-defined quantity (if periodic boundary is taken for the other two directions), which may be calculated. Suppose we are interested in the Chern number, $C$, at some Zeeman field $\Delta_Z=\Delta_0>0$. Then since time-reversal reverses the Chern number, we know for $\Delta_Z=-\Delta_0$, the Chern number must be $-C$. Consider a 3D space spanned by $q_{1,2}$ and $\Delta_Z$, then from Gauss's law, the Chern number change from $\Delta_Z=-\Delta_0$ to $\Delta_0$ equals the total monopole charge between these two planes in the 3D parameter space. The monopole, or gap closing point, is always at $(q_1,q_2,\Delta_Z)=0$, around which the Hamiltonian is that of 3D Weyl fermions: $h(q_1,q_2,q_3)=\sum_{i,j=1,2,3}A_{ij}\sigma_iq_j$, where $q_3\equiv{\Delta_Z}$. The charge of such a monopole is $\textrm{sign}\det(A)$, and since there are in total four such monopoles between $\Delta_Z=\pm\Delta_0$, we have the difference in Chern number $C-(-C)=4\textrm{sign}(\det{}A)$, or $C=2\textrm{sign}(\det{}A)$. All Chern numbers obtained in the text are derived using this method.

\subsection{Diagonalizing the Hamiltonian in equation (\ref{eq:hybridization})}
\label{sec:dispersion}
A Hamiltonian that describes isolated top and surface states around $D_i$ is
\bea
\tilde{H}_i=\left(\begin{matrix}
H^t_i & 0\\
0 & H^b_i
\end{matrix}\right),
\eea
and hybridization is equivalent to adding an off-diagonal block term, resulting in, to the lowest order in $|q|$,
\bea
\tilde{H}_i=\left(\begin{matrix}
H^t_i & \Delta_HI_{2\times2}\\
\Delta_HI_{2\times2} & H^b_i
\end{matrix}\right).
\eea
Diagonalizing $\tilde{H}_1$ directly, we obtain four bands:\begin{widetext}
\bea
E_1(\bq)&=&v_0q_1+\sqrt{{\Delta^t_1}^2+{\Delta^t_1}^2+2\Delta_H^2+2q_1^2v_1^2+2q_2^2v_2^2+\sqrt{(\Delta^t_1-\Delta^b_1)^2+4[(\Delta^t_1+\Delta^b_1)^2+4v_1^2q_1^2+4v_2^2q_2^2]}},\\
E_2(\bq)&=&v_0q_1+\sqrt{{\Delta^t_1}^2+{\Delta^t_1}^2+2\Delta_H^2+2q_1^2v_1^2+2q_2^2v_2^2-\sqrt{(\Delta^t_1-\Delta^b_1)^2+4[(\Delta^t_1+\Delta^b_1)^2+4v_1^2q_1^2+4v_2^2q_2^2]}},\\
E_3(\bq)&=&v_0q_1-\sqrt{{\Delta^t_1}^2+{\Delta^t_1}^2+2\Delta_H^2+2q_1^2v_1^2+2q_2^2v_2^2-\sqrt{(\Delta^t_1-\Delta^b_1)^2+4[(\Delta^t_1+\Delta^b_1)^2+4v_1^2q_1^2+4v_2^2q_2^2]}},\\
E_4(\bq)&=&v_0q_1-\sqrt{{\Delta^t_1}^2+{\Delta^t_1}^2+2\Delta_H^2+2q_1^2v_1^2+2q_2^2v_2^2+\sqrt{(\Delta^t_1-\Delta^b_1)^2+4[(\Delta^t_1+\Delta^b_1)^2+4v_1^2q_1^2+4v_2^2q_2^2]}}.
\eea\end{widetext}
Straightforward algebraic work shows that the \emph{only} solution for $E_2(\bq)=E_3(\bq)$, i.e., a gap-closing point, exists at $q_1=q_2=0$ when $|\Delta_H|=\sqrt{\Delta_1^t\Delta_1^b}$.

Parallel discussion for $D_{2,1',2'}$ proceeds and we conclude that a topological phase transition happens when 
\bea\label{eq:transition}
|\Delta_H|=\sqrt{\Delta^t_i\Delta_i^b},
\eea
whereas the Chern number contributed by the cone at $D_i$ changes from $\pm1$, depending on the sign of $\Delta_i^{t,b}$, to zero. Mark that on the right hand side of equation (\ref{eq:transition}), if $\Delta^t_i\Delta^b_i<0$, the transition cannot happen at any $\Delta_H$.

\textbf{Acknowledgements} CF and BAB thank A. Yazdani, R. J. Cava, N. P. Ong, and A. Alexandradinata for helpful discussions. CF specially thanks J. Liu and H. Lin for providing useful information on thin-film samples. CF is supported by ONR-N00014-11-1-0635. MJG acknowledges support from the AFOSR under grant FA9550-10-1-0459 and the ONR under grant N0014-11-1-0728. BAB was supported by NSF CAREER DMR- 095242, ONR-N00014-11-1-0635, Darpa- N66001-11-1-4110, David and Lucile Packard Foundation, and MURI-130-6082.


\begin{thebibliography}{34}
\expandafter\ifx\csname natexlab\endcsname\relax\def\natexlab#1{#1}\fi
\expandafter\ifx\csname bibnamefont\endcsname\relax
  \def\bibnamefont#1{#1}\fi
\expandafter\ifx\csname bibfnamefont\endcsname\relax
  \def\bibfnamefont#1{#1}\fi
\expandafter\ifx\csname citenamefont\endcsname\relax
  \def\citenamefont#1{#1}\fi
\expandafter\ifx\csname url\endcsname\relax
  \def\url#1{\texttt{#1}}\fi
\expandafter\ifx\csname urlprefix\endcsname\relax\def\urlprefix{URL }\fi
\providecommand{\bibinfo}[2]{#2}
\providecommand{\eprint}[2][]{\url{#2}}

\bibitem[{\citenamefont{Klitzing et~al.}(1980)\citenamefont{Klitzing, Dorda,
  and Pepper}}]{Klitzing}
\bibinfo{author}{\bibfnamefont{K.~v.} \bibnamefont{Klitzing}},
  \bibinfo{author}{\bibfnamefont{G.}~\bibnamefont{Dorda}}, \bibnamefont{and}
  \bibinfo{author}{\bibfnamefont{M.}~\bibnamefont{Pepper}},
  \bibinfo{journal}{Phys. Rev. Lett.} \textbf{\bibinfo{volume}{45}},
  \bibinfo{pages}{494} (\bibinfo{year}{1980}).

\bibitem[{\citenamefont{Haldane}(1988)}]{Haldane1988}
\bibinfo{author}{\bibfnamefont{F.~D.~M.} \bibnamefont{Haldane}},
  \bibinfo{journal}{Phys. Rev. Lett.} \textbf{\bibinfo{volume}{61}},
  \bibinfo{pages}{2015} (\bibinfo{year}{1988}),
  \urlprefix\url{http://link.aps.org/doi/10.1103/PhysRevLett.61.2015}.

\bibitem[{\citenamefont{Onoda and Nagaosa}(2003)}]{Onoda2003}
\bibinfo{author}{\bibfnamefont{M.}~\bibnamefont{Onoda}} \bibnamefont{and}
  \bibinfo{author}{\bibfnamefont{N.}~\bibnamefont{Nagaosa}},
  \bibinfo{journal}{Phys. Rev. Lett.} \textbf{\bibinfo{volume}{90}},
  \bibinfo{pages}{206601} (\bibinfo{year}{2003}),
  \urlprefix\url{http://link.aps.org/doi/10.1103/PhysRevLett.90.206601}.

\bibitem[{\citenamefont{Qi et~al.}(2008)\citenamefont{Qi, Hughes, and
  Zhang}}]{Qi2008}
\bibinfo{author}{\bibfnamefont{X.-L.} \bibnamefont{Qi}},
  \bibinfo{author}{\bibfnamefont{T.~L.} \bibnamefont{Hughes}},
  \bibnamefont{and} \bibinfo{author}{\bibfnamefont{S.-C.} \bibnamefont{Zhang}},
  \bibinfo{journal}{Phys. Rev. B} \textbf{\bibinfo{volume}{78}},
  \bibinfo{pages}{195424} (\bibinfo{year}{2008}),
  \urlprefix\url{http://link.aps.org/doi/10.1103/PhysRevB.78.195424}.

\bibitem[{\citenamefont{Yu et~al.}(2010)\citenamefont{Yu, Zhang, Zhang, Zhang,
  Dai, and Fang}}]{Yu2010}
\bibinfo{author}{\bibfnamefont{R.}~\bibnamefont{Yu}},
  \bibinfo{author}{\bibfnamefont{W.}~\bibnamefont{Zhang}},
  \bibinfo{author}{\bibfnamefont{H.-J.} \bibnamefont{Zhang}},
  \bibinfo{author}{\bibfnamefont{S.-C.} \bibnamefont{Zhang}},
  \bibinfo{author}{\bibfnamefont{X.}~\bibnamefont{Dai}}, \bibnamefont{and}
  \bibinfo{author}{\bibfnamefont{Z.}~\bibnamefont{Fang}},
  \bibinfo{journal}{Science} \textbf{\bibinfo{volume}{329}},
  \bibinfo{pages}{61} (\bibinfo{year}{2010}),
  \eprint{http://www.sciencemag.org/content/329/5987/61.full.pdf},
  \urlprefix\url{http://www.sciencemag.org/content/329/5987/61.abstract}.

\bibitem[{\citenamefont{Nomura and Nagaosa}(2011)}]{Nomura2011}
\bibinfo{author}{\bibfnamefont{K.}~\bibnamefont{Nomura}} \bibnamefont{and}
  \bibinfo{author}{\bibfnamefont{N.}~\bibnamefont{Nagaosa}},
  \bibinfo{journal}{Phys. Rev. Lett.} \textbf{\bibinfo{volume}{106}},
  \bibinfo{pages}{166802} (\bibinfo{year}{2011}),
  \urlprefix\url{http://link.aps.org/doi/10.1103/PhysRevLett.106.166802}.

\bibitem[{\citenamefont{Jiang et~al.}(2012)\citenamefont{Jiang, Qiao, Liu, and
  Niu}}]{Jiang2012}
\bibinfo{author}{\bibfnamefont{H.}~\bibnamefont{Jiang}},
  \bibinfo{author}{\bibfnamefont{Z.}~\bibnamefont{Qiao}},
  \bibinfo{author}{\bibfnamefont{H.}~\bibnamefont{Liu}}, \bibnamefont{and}
  \bibinfo{author}{\bibfnamefont{Q.}~\bibnamefont{Niu}},
  \bibinfo{journal}{Phys. Rev. B} \textbf{\bibinfo{volume}{85}},
  \bibinfo{pages}{045445} (\bibinfo{year}{2012}),
  \urlprefix\url{http://link.aps.org/doi/10.1103/PhysRevB.85.045445}.

\bibitem[{\citenamefont{Wang et~al.}(2013{\natexlab{a}})\citenamefont{Wang,
  Lian, Zhang, Xu, and Zhang}}]{Wang2013}
\bibinfo{author}{\bibfnamefont{J.}~\bibnamefont{Wang}},
  \bibinfo{author}{\bibfnamefont{B.}~\bibnamefont{Lian}},
  \bibinfo{author}{\bibfnamefont{H.}~\bibnamefont{Zhang}},
  \bibinfo{author}{\bibfnamefont{Y.}~\bibnamefont{Xu}}, \bibnamefont{and}
  \bibinfo{author}{\bibfnamefont{S.-C.} \bibnamefont{Zhang}},
  \bibinfo{journal}{arXiv:1305.7500v1}  (\bibinfo{year}{2013}{\natexlab{a}}).

\bibitem[{\citenamefont{Chang et~al.}(2013)\citenamefont{Chang, Zhang, Feng,
  Shen, Zhang, Guo, Li, Ou, Wei, Wang et~al.}}]{Chang2013}
\bibinfo{author}{\bibfnamefont{C.-Z.} \bibnamefont{Chang}},
  \bibinfo{author}{\bibfnamefont{J.}~\bibnamefont{Zhang}},
  \bibinfo{author}{\bibfnamefont{X.}~\bibnamefont{Feng}},
  \bibinfo{author}{\bibfnamefont{J.}~\bibnamefont{Shen}},
  \bibinfo{author}{\bibfnamefont{Z.}~\bibnamefont{Zhang}},
  \bibinfo{author}{\bibfnamefont{M.}~\bibnamefont{Guo}},
  \bibinfo{author}{\bibfnamefont{K.}~\bibnamefont{Li}},
  \bibinfo{author}{\bibfnamefont{Y.}~\bibnamefont{Ou}},
  \bibinfo{author}{\bibfnamefont{P.}~\bibnamefont{Wei}},
  \bibinfo{author}{\bibfnamefont{L.-L.} \bibnamefont{Wang}},
  \bibnamefont{et~al.}, \bibinfo{journal}{Science}
  \textbf{\bibinfo{volume}{340}}, \bibinfo{pages}{167} (\bibinfo{year}{2013}),
  \eprint{http://www.sciencemag.org/content/340/6129/167.full.pdf},
  \urlprefix\url{http://www.sciencemag.org/content/340/6129/167.abstract}.

\bibitem[{\citenamefont{Hasan and Kane}(2010)}]{hasan2010}
\bibinfo{author}{\bibfnamefont{M.~Z.} \bibnamefont{Hasan}} \bibnamefont{and}
  \bibinfo{author}{\bibfnamefont{C.~L.} \bibnamefont{Kane}},
  \bibinfo{journal}{Rev. Mod. Phys.} \textbf{\bibinfo{volume}{82}},
  \bibinfo{pages}{3045} (\bibinfo{year}{2010}).

\bibitem[{\citenamefont{Qi and Zhang}(2011)}]{qi2011rev}
\bibinfo{author}{\bibfnamefont{X.~L.} \bibnamefont{Qi}} \bibnamefont{and}
  \bibinfo{author}{\bibfnamefont{S.~C.} \bibnamefont{Zhang}},
  \bibinfo{journal}{Rev. Mod. Phys.} \textbf{\bibinfo{volume}{83}},
  \bibinfo{pages}{1057} (\bibinfo{year}{2011}).

\bibitem[{\citenamefont{Bernevig and Hughes}(2013)}]{Bernevig2013}
\bibinfo{author}{\bibfnamefont{A.~B.} \bibnamefont{Bernevig}} \bibnamefont{and}
  \bibinfo{author}{\bibfnamefont{T.~L.} \bibnamefont{Hughes}},
  \emph{\bibinfo{title}{Topological Insulators and Topological
  Superconductors}} (\bibinfo{publisher}{Princeton University Press},
  \bibinfo{year}{2013}).

\bibitem[{\citenamefont{Fu}(2011)}]{Fu2011}
\bibinfo{author}{\bibfnamefont{L.}~\bibnamefont{Fu}}, \bibinfo{journal}{Phys.
  Rev. Lett.} \textbf{\bibinfo{volume}{106}}, \bibinfo{pages}{106802}
  (\bibinfo{year}{2011}).

\bibitem[{\citenamefont{Hsieh et~al.}(2012)\citenamefont{Hsieh, Lin, Liu, Duan,
  Bansil, and Fu}}]{Hsieh2012}
\bibinfo{author}{\bibfnamefont{T.}~\bibnamefont{Hsieh}},
  \bibinfo{author}{\bibfnamefont{H.}~\bibnamefont{Lin}},
  \bibinfo{author}{\bibfnamefont{J.}~\bibnamefont{Liu}},
  \bibinfo{author}{\bibfnamefont{W.}~\bibnamefont{Duan}},
  \bibinfo{author}{\bibfnamefont{A.}~\bibnamefont{Bansil}}, \bibnamefont{and}
  \bibinfo{author}{\bibfnamefont{L.}~\bibnamefont{Fu}},
  \bibinfo{journal}{arXiv:1202.1003}  (\bibinfo{year}{2012}).

\bibitem[{\citenamefont{Xu et~al.}(2012)\citenamefont{Xu, Liu, Alidoust, Qian,
  Neupane, Denlinger, Wang, Wray, Cava, Lin et~al.}}]{Xu2012}
\bibinfo{author}{\bibfnamefont{S.-Y.} \bibnamefont{Xu}},
  \bibinfo{author}{\bibfnamefont{C.}~\bibnamefont{Liu}},
  \bibinfo{author}{\bibfnamefont{N.}~\bibnamefont{Alidoust}},
  \bibinfo{author}{\bibfnamefont{D.}~\bibnamefont{Qian}},
  \bibinfo{author}{\bibfnamefont{M.}~\bibnamefont{Neupane}},
  \bibinfo{author}{\bibfnamefont{J.~D.} \bibnamefont{Denlinger}},
  \bibinfo{author}{\bibfnamefont{Y.~J.} \bibnamefont{Wang}},
  \bibinfo{author}{\bibfnamefont{L.~A.} \bibnamefont{Wray}},
  \bibinfo{author}{\bibfnamefont{R.~J.} \bibnamefont{Cava}},
  \bibinfo{author}{\bibfnamefont{H.}~\bibnamefont{Lin}}, \bibnamefont{et~al.},
  \bibinfo{journal}{Nat Commu} \textbf{\bibinfo{volume}{3}},
  \bibinfo{pages}{1192} (\bibinfo{year}{2012}).

\bibitem[{\citenamefont{Dziawa et~al.}(2012)\citenamefont{Dziawa, Kowalski,
  Dybko, Buczko, Szczerbakow, Szot, Ausakowska, Balasubramanian, Wojek,
  Berntsen et~al.}}]{Dziawa2012}
\bibinfo{author}{\bibfnamefont{P.}~\bibnamefont{Dziawa}},
  \bibinfo{author}{\bibfnamefont{B.~J.} \bibnamefont{Kowalski}},
  \bibinfo{author}{\bibfnamefont{K.}~\bibnamefont{Dybko}},
  \bibinfo{author}{\bibfnamefont{R.}~\bibnamefont{Buczko}},
  \bibinfo{author}{\bibfnamefont{A.}~\bibnamefont{Szczerbakow}},
  \bibinfo{author}{\bibfnamefont{M.}~\bibnamefont{Szot}},
  \bibinfo{author}{\bibfnamefont{E.}~\bibnamefont{Ausakowska}},
  \bibinfo{author}{\bibfnamefont{T.}~\bibnamefont{Balasubramanian}},
  \bibinfo{author}{\bibfnamefont{B.~M.} \bibnamefont{Wojek}},
  \bibinfo{author}{\bibfnamefont{M.~H.} \bibnamefont{Berntsen}},
  \bibnamefont{et~al.}, \bibinfo{journal}{Nature Materials}
  \textbf{\bibinfo{volume}{advance online publication}} (\bibinfo{year}{2012}).

\bibitem[{\citenamefont{Tanaka et~al.}(2012)\citenamefont{Tanaka, Ren, Sato,
  Nakayama, Souma, Takahashi, Segawa, and Ando}}]{Tanaka2012}
\bibinfo{author}{\bibfnamefont{Y.}~\bibnamefont{Tanaka}},
  \bibinfo{author}{\bibfnamefont{Z.}~\bibnamefont{Ren}},
  \bibinfo{author}{\bibfnamefont{T.}~\bibnamefont{Sato}},
  \bibinfo{author}{\bibfnamefont{K.}~\bibnamefont{Nakayama}},
  \bibinfo{author}{\bibfnamefont{S.}~\bibnamefont{Souma}},
  \bibinfo{author}{\bibfnamefont{T.}~\bibnamefont{Takahashi}},
  \bibinfo{author}{\bibfnamefont{K.}~\bibnamefont{Segawa}}, \bibnamefont{and}
  \bibinfo{author}{\bibfnamefont{Y.}~\bibnamefont{Ando}},
  \bibinfo{journal}{Nat. Phys.} \textbf{\bibinfo{volume}{8}},
  \bibinfo{pages}{800} (\bibinfo{year}{2012}).

\bibitem[{\citenamefont{Fang et~al.}(2012)\citenamefont{Fang, Gilbert, Xu,
  Bernevig, and Hasan}}]{Fang2012a}
\bibinfo{author}{\bibfnamefont{C.}~\bibnamefont{Fang}},
  \bibinfo{author}{\bibfnamefont{M.~J.} \bibnamefont{Gilbert}},
  \bibinfo{author}{\bibfnamefont{S.-Y.} \bibnamefont{Xu}},
  \bibinfo{author}{\bibfnamefont{B.~A.} \bibnamefont{Bernevig}},
  \bibnamefont{and} \bibinfo{author}{\bibfnamefont{M.~Z.} \bibnamefont{Hasan}},
  \bibinfo{journal}{arXiv:1212.3285v1}  (\bibinfo{year}{2012}).

\bibitem[{\citenamefont{Wang et~al.}(2013{\natexlab{b}})\citenamefont{Wang,
  Tsai, Lin, Xu, Neupane, Hasan, and Bansil}}]{Wang2013a}
\bibinfo{author}{\bibfnamefont{Y.~J.} \bibnamefont{Wang}},
  \bibinfo{author}{\bibfnamefont{W.-F.} \bibnamefont{Tsai}},
  \bibinfo{author}{\bibfnamefont{H.}~\bibnamefont{Lin}},
  \bibinfo{author}{\bibfnamefont{S.-Y.} \bibnamefont{Xu}},
  \bibinfo{author}{\bibfnamefont{M.}~\bibnamefont{Neupane}},
  \bibinfo{author}{\bibfnamefont{M.}~\bibnamefont{Hasan}}, \bibnamefont{and}
  \bibinfo{author}{\bibfnamefont{A.}~\bibnamefont{Bansil}},
  \bibinfo{journal}{arXiv:1304.8119}  (\bibinfo{year}{2013}{\natexlab{b}}).

\bibitem[{\citenamefont{Liu et~al.}(2013{\natexlab{a}})\citenamefont{Liu, Duan,
  and Fu}}]{Liu2013}
\bibinfo{author}{\bibfnamefont{J.}~\bibnamefont{Liu}},
  \bibinfo{author}{\bibfnamefont{W.}~\bibnamefont{Duan}}, \bibnamefont{and}
  \bibinfo{author}{\bibfnamefont{L.}~\bibnamefont{Fu}},
  \bibinfo{journal}{arXiv:1304.0430}  (\bibinfo{year}{2013}{\natexlab{a}}).

\bibitem[{\citenamefont{Okada et~al.}(2013)\citenamefont{Okada, Serbyn, Lin,
  Walkup, Zhou, Dhital, Neupane, Xu, Wang, Sankar et~al.}}]{Okada2013}
\bibinfo{author}{\bibfnamefont{Y.}~\bibnamefont{Okada}},
  \bibinfo{author}{\bibfnamefont{M.}~\bibnamefont{Serbyn}},
  \bibinfo{author}{\bibfnamefont{H.}~\bibnamefont{Lin}},
  \bibinfo{author}{\bibfnamefont{D.}~\bibnamefont{Walkup}},
  \bibinfo{author}{\bibfnamefont{W.}~\bibnamefont{Zhou}},
  \bibinfo{author}{\bibfnamefont{C.}~\bibnamefont{Dhital}},
  \bibinfo{author}{\bibfnamefont{M.}~\bibnamefont{Neupane}},
  \bibinfo{author}{\bibfnamefont{S.}~\bibnamefont{Xu}},
  \bibinfo{author}{\bibfnamefont{Y.~J.} \bibnamefont{Wang}},
  \bibinfo{author}{\bibfnamefont{R.}~\bibnamefont{Sankar}},
  \bibnamefont{et~al.}, \bibinfo{journal}{arXiv:1305.2823}
  (\bibinfo{year}{2013}).

\bibitem[{\citenamefont{Elleman and Wilman}(1948)}]{Elleman1948}
\bibinfo{author}{\bibfnamefont{A.~J.} \bibnamefont{Elleman}} \bibnamefont{and}
  \bibinfo{author}{\bibfnamefont{H.}~\bibnamefont{Wilman}},
  \bibinfo{journal}{Proceedings of the Physical Society}
  \textbf{\bibinfo{volume}{61}}, \bibinfo{pages}{164} (\bibinfo{year}{1948}),
  \urlprefix\url{http://stacks.iop.org/0959-5309/61/i=2/a=307}.

\bibitem[{\citenamefont{Bylander}(1966)}]{Bylander1966}
\bibinfo{author}{\bibfnamefont{E.~G.} \bibnamefont{Bylander}},
  \bibinfo{journal}{Mater. Sci. Eng.} \textbf{\bibinfo{volume}{1}},
  \bibinfo{pages}{190} (\bibinfo{year}{1966}).

\bibitem[{\citenamefont{Taskin et~al.}(2013)\citenamefont{Taskin, Sasaki,
  Segawa, and Ando}}]{Taskin2013}
\bibinfo{author}{\bibfnamefont{A.~A.} \bibnamefont{Taskin}},
  \bibinfo{author}{\bibfnamefont{S.}~\bibnamefont{Sasaki}},
  \bibinfo{author}{\bibfnamefont{K.}~\bibnamefont{Segawa}}, \bibnamefont{and}
  \bibinfo{author}{\bibfnamefont{Y.}~\bibnamefont{Ando}},
  \bibinfo{journal}{arXiv:1305.2470}  (\bibinfo{year}{2013}).

\bibitem[{\citenamefont{Mathur et~al.}(1970)\citenamefont{Mathur, Deis, Jones,
  Patterson, and et~al.}}]{Mathur1970}
\bibinfo{author}{\bibfnamefont{M.~P.} \bibnamefont{Mathur}},
  \bibinfo{author}{\bibfnamefont{D.~W.} \bibnamefont{Deis}},
  \bibinfo{author}{\bibfnamefont{C.~K.} \bibnamefont{Jones}},
  \bibinfo{author}{\bibfnamefont{A.}~\bibnamefont{Patterson}},
  \bibnamefont{and} \bibinfo{author}{\bibfnamefont{W.~J.~C.}
  \bibnamefont{et~al.}}, \bibinfo{journal}{J. Appl. Phys.}
  \textbf{\bibinfo{volume}{41}}, \bibinfo{pages}{1005} (\bibinfo{year}{1970}).

\bibitem[{\citenamefont{Inoue et~al.}(1976)\citenamefont{Inoue, Yagi, Ishii,
  and Tatsukawa}}]{Inoue1976}
\bibinfo{author}{\bibfnamefont{M.}~\bibnamefont{Inoue}},
  \bibinfo{author}{\bibfnamefont{H.}~\bibnamefont{Yagi}},
  \bibinfo{author}{\bibfnamefont{K.}~\bibnamefont{Ishii}}, \bibnamefont{and}
  \bibinfo{author}{\bibfnamefont{T.}~\bibnamefont{Tatsukawa}},
  \bibinfo{journal}{J. Lo. Temp. Phys.} \textbf{\bibinfo{volume}{23}},
  \bibinfo{pages}{785} (\bibinfo{year}{1976}).

\bibitem[{\citenamefont{Nielsen et~al.}(2012)\citenamefont{Nielsen, Levin,
  Jaworski, Schmidt-Rohr, and Heremans}}]{Nielsen2012}
\bibinfo{author}{\bibfnamefont{M.~D.} \bibnamefont{Nielsen}},
  \bibinfo{author}{\bibfnamefont{E.~M.} \bibnamefont{Levin}},
  \bibinfo{author}{\bibfnamefont{C.~M.} \bibnamefont{Jaworski}},
  \bibinfo{author}{\bibfnamefont{K.}~\bibnamefont{Schmidt-Rohr}},
  \bibnamefont{and} \bibinfo{author}{\bibfnamefont{J.~P.}
  \bibnamefont{Heremans}}, \bibinfo{journal}{Phys. Rev. B}
  \textbf{\bibinfo{volume}{85}}, \bibinfo{pages}{045210}
  (\bibinfo{year}{2012}),
  \urlprefix\url{http://link.aps.org/doi/10.1103/PhysRevB.85.045210}.

\bibitem[{\citenamefont{Berlincourt and Jaffe}(1958)}]{Berlincourt1958}
\bibinfo{author}{\bibfnamefont{D.}~\bibnamefont{Berlincourt}} \bibnamefont{and}
  \bibinfo{author}{\bibfnamefont{H.}~\bibnamefont{Jaffe}},
  \bibinfo{journal}{Phys. Rev.} \textbf{\bibinfo{volume}{111}},
  \bibinfo{pages}{143} (\bibinfo{year}{1958}),
  \urlprefix\url{http://link.aps.org/doi/10.1103/PhysRev.111.143}.

\bibitem[{\citenamefont{Bierly et~al.}(1963)\citenamefont{Bierly, Muldawer, and
  Beckman}}]{Bierly1963}
\bibinfo{author}{\bibfnamefont{J.}~\bibnamefont{Bierly}},
  \bibinfo{author}{\bibfnamefont{L.}~\bibnamefont{Muldawer}}, \bibnamefont{and}
  \bibinfo{author}{\bibfnamefont{O.}~\bibnamefont{Beckman}},
  \bibinfo{journal}{Acta Metallurgica} \textbf{\bibinfo{volume}{11}},
  \bibinfo{pages}{447 } (\bibinfo{year}{1963}), ISSN \bibinfo{issn}{0001-6160},
  \urlprefix\url{http://www.sciencedirect.com/science/article/pii/0001616063901706}.

\bibitem[{\citenamefont{Littlewood et~al.}(2010)\citenamefont{Littlewood,
  Mihaila, Schulze, Safarik, Gubernatis, Bostwick, Rotenberg, Opeil,
  Durakiewicz, Smith et~al.}}]{Littlewood2010}
\bibinfo{author}{\bibfnamefont{P.~B.} \bibnamefont{Littlewood}},
  \bibinfo{author}{\bibfnamefont{B.}~\bibnamefont{Mihaila}},
  \bibinfo{author}{\bibfnamefont{R.~K.} \bibnamefont{Schulze}},
  \bibinfo{author}{\bibfnamefont{D.~J.} \bibnamefont{Safarik}},
  \bibinfo{author}{\bibfnamefont{J.~E.} \bibnamefont{Gubernatis}},
  \bibinfo{author}{\bibfnamefont{A.}~\bibnamefont{Bostwick}},
  \bibinfo{author}{\bibfnamefont{E.}~\bibnamefont{Rotenberg}},
  \bibinfo{author}{\bibfnamefont{C.~P.} \bibnamefont{Opeil}},
  \bibinfo{author}{\bibfnamefont{T.}~\bibnamefont{Durakiewicz}},
  \bibinfo{author}{\bibfnamefont{J.~L.} \bibnamefont{Smith}},
  \bibnamefont{et~al.}, \bibinfo{journal}{Phys. Rev. Lett.}
  \textbf{\bibinfo{volume}{105}}, \bibinfo{pages}{086404}
  (\bibinfo{year}{2010}),
  \urlprefix\url{http://link.aps.org/doi/10.1103/PhysRevLett.105.086404}.

\bibitem[{\citenamefont{Lin}(2013)}]{Lin}
\bibinfo{author}{\bibfnamefont{H.}~\bibnamefont{Lin}},
  \bibinfo{journal}{Private communications}  (\bibinfo{year}{2013}).

\bibitem[{\citenamefont{Liu et~al.}(2013{\natexlab{b}})\citenamefont{Liu,
  Hsieh, Duan, Moodera, and Fu}}]{Liu2013b}
\bibinfo{author}{\bibfnamefont{J.}~\bibnamefont{Liu}},
  \bibinfo{author}{\bibfnamefont{T.~H.} \bibnamefont{Hsieh}},
  \bibinfo{author}{\bibfnamefont{W.}~\bibnamefont{Duan}},
  \bibinfo{author}{\bibfnamefont{J.}~\bibnamefont{Moodera}}, \bibnamefont{and}
  \bibinfo{author}{\bibfnamefont{L.}~\bibnamefont{Fu}}, in
  \emph{\bibinfo{booktitle}{APS March Meeting}}
  (\bibinfo{year}{2013}{\natexlab{b}}).

\bibitem[{\citenamefont{\textrm{X.L. Qi}
  et~al.}(2006)\citenamefont{\textrm{X.L. Qi}, \textrm{Y.S. Wu}, and
  \textrm{S.C. Zhang}}}]{qi2006}
\bibinfo{author}{\bibnamefont{\textrm{X.L. Qi}}},
  \bibinfo{author}{\bibnamefont{\textrm{Y.S. Wu}}}, \bibnamefont{and}
  \bibinfo{author}{\bibnamefont{\textrm{S.C. Zhang}}}, \bibinfo{journal}{Phys.
  Rev. B} \textbf{\bibinfo{volume}{74}}, \bibinfo{pages}{045125}
  (\bibinfo{year}{2006}).

\bibitem[{\citenamefont{Kohmoto}(1985)}]{Kohmoto1985}
\bibinfo{author}{\bibfnamefont{M.}~\bibnamefont{Kohmoto}},
  \bibinfo{journal}{Annals of Physics} \textbf{\bibinfo{volume}{160}},
  \bibinfo{pages}{343 } (\bibinfo{year}{1985}), ISSN \bibinfo{issn}{0003-4916},
  \urlprefix\url{http://www.sciencedirect.com/science/article/pii/0003491685901484}.

\end{thebibliography}

\end{document}